# Consequences of broken axial symmetry in heavy nuclei – an overview of the situation in the valley of stability.

E. Grosse, A. R. Junghans and J. N. Wilson

Institute of Nuclear and Particle Physics, Technische Universität 01062 Dresden, Germany,

Institute of Radiation Physics, Helmholtz-Zentrum Dresden-Rossendorf, 01314 Dresden, Germany

and

Institute of Nuclear Physics, Université Paris-Sud and CNRS/IN2P3, 91406 Orsay, France

**Abstract:** An overview on the various effects of axial symmetry breaking is presented for medium heavy and heavy nuclei covering the mass number range 70 < A <240. The discussion includes various observations for nuclei: level densities, spectroscopic features as energies and transition rates, ground state masses and finally the splitting of giant dipole resonances. Quadrupole moments and rates can be derived from models of triaxial rigid rotation or cranking for a given triaxiality parameters γ, but microscopic considerations are needed to predict these for each nucleus investigated. Respective predictions were made by recently made Hartree-Fock-Bogolyubov (HFB) calculations extended to arbitrary triaxiality by a generator coordinate method. In accord to these, various observations as reported in this overview demonstrate the importance of allowing a breaking of axial symmetry for heavy nuclei already in the valley of stability. Considering this breaking as indicated from the HFB approach surprisingly many experimental data are well described globally without the need for local fit parameters. In addition to these comparisons it will be shown that it is advantageous to consider $c_\gamma=\cos(3\gamma)$ an indicator of axiality for heavy nuclei independent of their quadrupole moment.

## I.  Introduction

The detection of large quadrupole moments $Q_s$ in heavy odd nuclei by hyperfine structure measurements of atomic transitions [1] has played an important role for the understanding of the structure of heavy nuclei. The departure of these moments from predictions by shell model considerations led to the postulation of broken spherical symmetry [2-4] for odd but also for even-even nuclei. The concept of an axially deformed core was introduced [5] for odd nuclei with strong deformation, *i.e.* those with large spectroscopic quadrupole moment $Q_s$, and complex particle shell model calculations based on it were presented for lanthanide and actinide nuclei. Based on the assumption of axial symmetry of the neighboring even nuclei the harmonic oscillator shell model was extended by the introduction of two frequencies per major shell, one for an oscillation perpendicular to the symmetry axis and one for displacements along it. This concept was extended in a phenomenological way to less deformed nuclei including a graphical continuation [5] versus the doubly magic $^{208}$Pb, assumed to be spherically symmetric. This increase of $Q_s$ in odd nuclei exceeding shell model predictions was described as being related to the intrinsic quadrupole moment $Q_i$ of the core nucleus which is also observable by Coulomb excitation [7] to its first excited $2^+$ state and enhanced E2-transitions [6] from them to ground. The observation of more than one $2^+$ state with enhanced electromagnetic quadrupole transition strength to the ground state inspired two alternative explanations:
1. the breaking of axial symmetry with a tensorial intrinsic quadrupole moment with different components in the two transverse directions, possibly indicating [8] static triaxiality.
2. semi-classical vibrations along or perpendicular to the symmetry axis (β- or γ- vibration), and thus proposing a more dynamic triaxiality [9].

The aim of this overview paper is to list various experimental observables especially related to the breaking of nuclear shape axiality and to outline their respective importance concerning these two alternatives. We will show that a manifestation of triaxiality may become obvious from nuclear level densities, in the spectroscopy of odd nuclei, in Coulomb excitation and subsequent reorientation of levels in even nuclei and their gamma decay as well as in radiative capture and photon absorption, especially via the isovector giant dipole resonances (IVGDR). In many even nuclei the splitting of the IVGDR had already indicated the role of nuclear deformation and hence this observable allows a good test for the breaking of axial symmetry [10, 11]. These and most of other experimental studies concerning an evidence for broken axial symmetry can be performed only on nuclei in or close to the valley of stability against beta-decay, and we will present an overview related to these. But in environments responsible for the cosmic nuclear synthesis some of the listed processes occur in exotic nuclei which are accessible to experiment not as easily as stable isotopes. Similarly, the knowledge of radiative neutron capture cross sections in actinide and other unstable nuclei is of central importance for the detailed understanding of nuclear power generation by fission and the simultaneous production of radionuclides, which have – at least partially – to be stored as eventually long lived radioactive waste, unless they are transmuted such that they decay faster. Thus it is important to find means for an extrapolation of the results of our study nuclei away from stability – especially concerning radiative capture cross sections. In addition to the discussion of the experimental situation we will hence give a short overview on related theoretical work.

## II. Nuclear level densities and the symmetry of excitation modes

The number of excited states per energy interval – and hence the state density – in a system of A nucleons depends on its total kinetic energy *E* and on the density of single particle states, respectively on the average Fermi momentum $k_F$ of the A nucleons within the nuclear volume. The state density $\omega_{qp}(U)$ in the intrinsic system can be derived from a combinatorial treatment of shell model results or approximated as a 2-component Fermi gas; here *U* is the excitation energy after subtraction of a shell correction term as known from a comparison of ground state masses to a liquid drop prediction [12]. In case only single particle states are regarded in a spherical nucleus a Gaussian shape with width σ can be assumed for the distribution of their M-values centred at M=0. The resulting density of quasi-particle levels with spin *I* near $E_x$, observable in the laboratory system, can be approximated [13 -17] by

$$\rho_{qp}(E_x, I) = (2I+1) \frac{e^{\frac{-I(I+1)}{2\sigma^2}}}{\sqrt{8\pi}\,\sigma^3} \omega_{qp}(U) \qquad (1).$$

Using σ from a Thomas-Fermi approach [14] the exponential spin cut off in Eq. (1) becomes important for *I*≥2 and causes a spin dependent peaking near angular momentum *I* = 4.

*Collective enhancement*

It has been shown [15 -20] that Eq. (1) underestimates experimental level densities for low spins by up to two orders of magnitude, if ω(*U*) is derived by combinatorial considerations or taken from a Fermi gas approach. An increase of the level density in heavy nuclei by the inclusion of their collective degrees of freedom was predicted [16, 18, 21- 23] and observed in several experiments as well. Most of the calculations follow a proposal for a rotational enhancement resulting from the build-up of a rotational band on each intrinsic quasi-particle state [21]; no mention of broken axial symmetry as possible cause for further enhancement can be found, even in work which still indicated a deficit as compared to observations [18]. In a general view on the effect of rotation presented at the Rochester symposium on fission [17] it was pointed out, that the collective level density enhancement is clearly larger, if the restriction to axial symmetry is released. Such an enhancement [19] was predicted to reach up to 30 MeV [24, 25] but a need for further study was proposed by an experimental study [23]. In view of the IVGDR splitting observed [26, 11] in nearly all nuclei with A>70 – also discussed later as well as other indicators for broken axiality – it is important to account for an equilibrium shape that possesses all the rotational degrees of freedom of a three-dimensional body. It thus violates rotational symmetry, in the sense that it is not invariant with respect to any rotation of the coordinate axes, such a rotational band on top of every intrinsic state with total collective angular momentum *I*, induces (2*I*+1) levels. *"Each of these levels is itself (2I+ l)-fold degenerate, corresponding to the different components M"* [17]. The rotation induced collective enhancement may fade out at higher energy, but this will happen only far above the energy range around $S_n$, of interest here [24, 23, 27]. An adiabatic approximation [17, 24, 28 - 30] for low spin shows that the proportionality between level density $\rho(E_x,I)$ and the density of the intrinsic states $\omega_{qp}(E)$ increases for non-symmetric shapes by $\sqrt{(8\pi)}\,\sigma^3/4 \gg 40$ as compared to neglecting collective rotation.

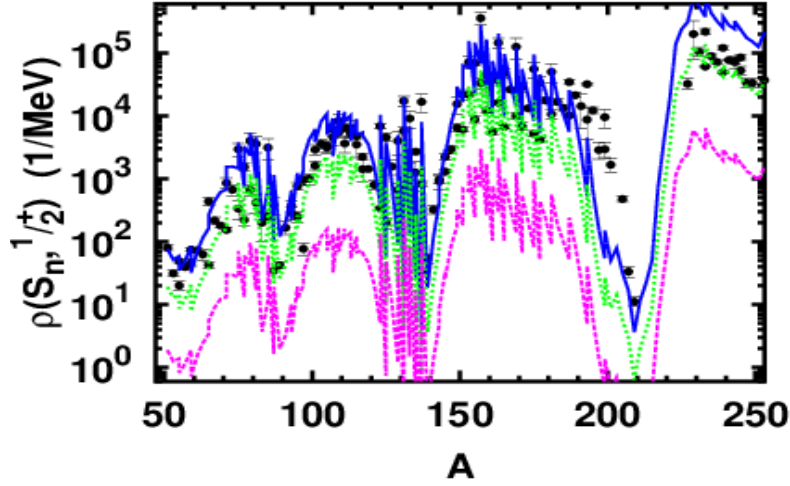

*Fig. 1*: Average level densities $\rho(S_n, \frac{1}{2}^+)$ in nuclei with 71<A<253 as observed in neutron capture by spin 0 target nuclei. Data (black dots Ɨ) compiled by the RIPL collaboration [31, 20] are compared to the prediction to be discussed in the next section; the curves depict the increase of $\rho$ in going from a situation without rotational enhancement (lowest), the situation of axiality (medium) to the enhancement caused by giving up spherical and axial symmetry (highest curve).

For other symmetry classes $\rho(E_x,I)$ is further modified [24] and in Fig. 1 the effect of various symmetry assumptions is depicted for spin ½ states in nuclei reached in n-capture by stable targets. An additional collective enhancement of $\rho(E_x,I)$ possibly due to shape oscillations and vibrational bands have been shown [17, 23, 33] to be rather small in heavy nuclei and are neglected here. When higher spin states correspond to a higher yrast-energy $E_{yr}$ a spin cut-off term reduces the enhanced $\rho(E,I)$ for increased $I$ :

$$\rho(E,I) = (2I+1)/4 \cdot \omega_{qp}(U - E_{yr}) \simeq (2I+1)/4 \cdot \omega_{qp}(U) \cdot e^{-E_{yr}(I)/T_{eff}} \qquad (2).$$

$T_{eff}$ denotes a spectral temperature, the inverse of the logarithmic derivative of $\rho(E,I)$. As demonstrated [30], the derivation of the total level density

$$\rho(E) = \sum_I \rho(E,I) \qquad (3)$$

needs complex considerations to quantify $E_{yr}$ as well as its dependence on the angular momentum $I$. It has become customary to describe this by an effective moment of inertia $\mathcal{J}$ in resemblance to a strongly deformed axial rotor. As recently indicated [34] by Hartree-Fock-Bogolyubov (HFB) calculations performed for a large number of nuclei the triaxial deformation parameter $\gamma$ may differ considerably from 0 (axiality). A better value for $E_{yr}$ and consequently $\rho(E)$ may be reached by allowing possible modifications away from the I(I+1) rule, derived [3] for axial rotors; it was shown [82] long ago to be valid only under the assumption of axiality. To estimate the effects due to the influence of γ on $E_{yr}$ an approximation for $E_{yr}$ was proposed by us [30], which we use here with minor modification:

$$E_{yr} \approx \hbar^2 \cdot (I + c_\gamma^2 I^2) / (2\mathcal{J}) \quad \text{with} \quad c_\gamma = \cos(3\gamma) \quad \text{and} \quad \mathcal{J}/\hbar^2 \approx \max(1.5, 77 \cdot c_\gamma^2) / MeV \qquad (4)$$

Such an estimate uses $c_\gamma$ as a kind of axiality-indicator; smooth changes for total level density predictions are expected as it does not affect the symmetry class. Data for elevated $I$ are shown in Fig. 6; results for $\rho(E_x \approx E_{yr}, I)$ were obtained in charged particle compound reactions [35] and a reasonable agreement to Eqs. (1-4) was found [30] for A=92 and A=238. Levels of smaller $I$ may be populated with neutrons and for these the rotational energy $E_{yr}$ and the exponential factor in Eq. (2) can be neglected. Unfortunately, respective experiments [36] do not allow to derive level densities in absolute units, but they show the interesting feature of a break in their logarithmic slope versus excitation. This can be related to a phase change by pairing condensation [37, 30] between a constant temperature regime to a Fermi gas. Such a combined model for level density predictions was published [15] long ago and a similar combination was compared to experiments on absolute scale recently [28, 29] and some results will be discussed in the next section.

### *Intrinsic state densities*

The typical mass density of heavy nuclei corresponds to a Fermi momentum of $k_F$ = 263 MeV [19]. The corresponding density of single particle states is $g(\varepsilon) = 6 \cdot \tilde{a} / \pi^2$ and a value $\tilde{a} \approx A/15$ was derived [19] in case of a homogeneous nucleon distribution in space. In any Fermionic system a pairing condensation sets in at a critical temperature $t_{pt}$ below which a transition to a paired Bosonic phase takes place [38]. In non-nuclear Fermionic systems a second order phase transition at a critical temperature is known to reduce the entropy and to change the partition function. Such a condensation into pairs, i.e. Bosons, causing a 2$^{nd}$ order phase transition with a change of entropy was also expected for nuclei [21, 37, 38] and for them this is likely to cause a break in the slope of $\omega_{qp}(E_x)$ versus $E_x$. In micro-canonical formalism based on the BCS theory [37] the sudden change in slope was predicted to be washed out, but in that work neither shell corrections nor a breaking of spherical symmetry are included. The Fermi gas transition temperature $t_{pt}$ = 0.567·$\Delta_o$ is A-dependent by its relation to the pairing gap, which is normally assumed to be given by $\Delta_o \approx 12 \cdot A^{-1/2}$. An approximation of the respective energy by $U_{pt} = \tilde{a} \cdot t_{pt}^2$ was shown [37] to be justified and it can be assumed that the Fermi gas expression Eq.(5) holds at and above this phase transition energy.

To apply these facts to a prediction of level densities in finite heavy nuclei additional points have to be regarded. Rotational motion is accounted for in Eq. (2), but the excitation energy $E_x$ has to be corrected for shell and pairing effects. Here a back-shift has been discussed since long [15, 39], and it can be set to the difference between the experimental ground state mass and the liquid drop model prediction [12] for it. By using $U = E_{FG} = E_x - E_{ld}$ one adapts $E_{ld}$ as the zero energy of the Fermi gas. Applying a saddle point approximation and taking equidistant shell model levels the quasi-particle state density for a 2-component Fermi gas in the intrinsic frame is [15, 19, 40]

$$\omega_{qp}(U) \approx \frac{\sqrt{\pi} e^{2\sqrt{\tilde{a}U}}}{12 \tilde{a}^{1/4} U^{5/4}} \qquad (5).$$

Recent reviews [20, 41] on the level density in heavy nuclei list back-shifted Fermi-gas calculations [19, 22, 23] as well as a combinatorial treatment of the states predicted in a nuclear shell model. The Fermi gas expression (5) can be considered an analytical approximation to results eventually arising from a microscopic approach. For energies below $U_{pt} = E_{pt} - E_{ld}$ new experimental data [35] favor an exponential extrapolation towards the ground state region, and this limit can well be estimated by $\rho_{gs} \approx 1/\Delta_o$. Such a constant temperature model was already compared to measurements previously [15, 20, 33, 42, 43], partially in combination to a Fermi gas model (FGM), and a deficit to experimental data is mentioned rather often. These studies have compensated the deficit by using $\tilde{a}$ as fit parameter [39, 40, 43] and $\tilde{a} \geq$ A/10 was favored, which clearly exceeds the nuclear matter value. The use of $\tilde{a}$ in the exponent in Eq. (5) has a large effect near $S_n$, and seemingly this suffices to come close to the observations. But it causes a very significant increase at higher energy and this is depicted in Fig. 2 for the density of $I^\pi$ = ½$^+$ levels in $^{97}$Mo. The clear disagreement shown there to a combinatorial calculation [25] supports the preference for non-axial rotational enhancement as proper means to increase the state density versus an increase of $\tilde{a}$ in predictions by the analytical Fermi gas approach; the latter proposition was often made [16, 19, 20, 22, 23, 39, 42] under the assumption of axiality. Hence the widespread habit to modify the level density parameter $\tilde{a}$ – proposed as phenomenological incorporation of shell effects [22] – or even to take it as a free local fit parameter is avoided now and this suppresses any mutual interference between the dependence of $\omega_{qp}(U)$ on A and $E_x$.

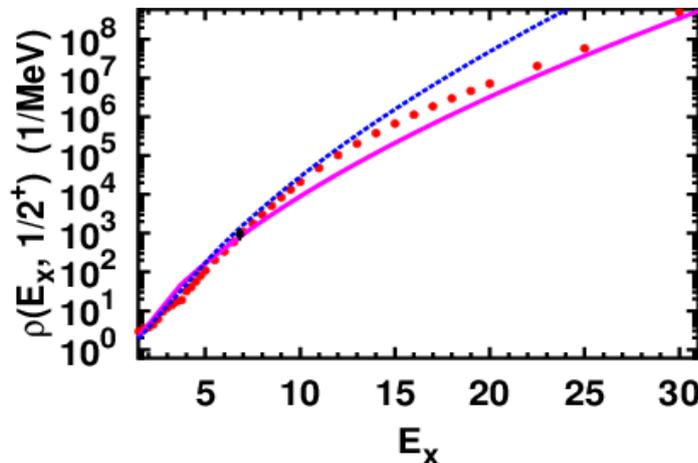

*Fig. 2: Average level densities $\rho(E_x,½^+)$ in $^{97}$Mo in dependence of excitation energy. An experimental data point taken near $S_n$ is shown (●) together with results of three calculations: Red dots (●) depict HFB plus combinatorial calculations [25, 20], the drawn line in magenta results from the calculation based on the FGM [28], which is explained in the text, and the dashed line in blue shows the change when assuming axiality together with an increase of the "level density parameter" $\tilde{a}$ from A/15 to A/10 and adjusting the absolute value at $S_n$ to experimental data.*

The advantage of the present FGM to a procedure with one based on an adjusted ã is obvious in Fig. 2 and from the agreement on absolute scale depicted in Fig. 1, which was prepared without any parameter fitting. This leads to the conclusion that for heavy nuclei in the valley of stability the assumption of axiality has to be reassessed before making arguments based on it. Other nuclear properties of relevance to that point will be discussed in the following.

### III. Nuclear spectroscopy and indications of broken axial symmetry

*Spectroscopic features of odd nuclei*

The electromagnetic response of nuclei has played an important role for the exploration of the size of nuclei and the departure of their shape from spherical symmetry: Hyperfine structure measurements of atomic transitions have delivered the first experimental manifestation of nuclear shapes [1, 44, 45]. In nearly 800 odd nuclei the 'spectroscopic' electric quadrupole moment $Q_S$ of the ground state was determined, mainly by accurate study of the splitting of atomic transitions – i.e. their hyperfine structure. In general the negative signs are clearly more frequent and the sizes of these quadrupole moments were often not in agreement to single particle or hole predictions.[19, 44]. In experiments with un-polarized probes only the quadrupole moment can be determined, as then atomic hyperfine structure is not sensitive to a possible non-axiality of the nuclear shape. In most publications on nuclear deformation the possibility of a broken axial symmetry is not mentioned, but early theoretical work on a deformed nuclear shell model [5] is explicitly dedicated to strongly deformed nuclei, for which axial symmetry is approximately conserved – as will also be discussed later. A subsequent paper [46] deals with the possible distortion due to an interplay between nucleonic motion and the inner nuclear core. For less deformed nuclei such a coupling to a core with broken axial symmetry was described by tilted axis cranking [47] and as an excitation mode with a "wobbling" induced by the core triaxiality. Such extra modes as well as "chiral" or "parallel" bands and "signature splitting" [48, 49] support the idea of increasing level densities by collective enhancement as discussed above. Experiments on the size of electric quadrupole (BE2) transitions [50 - 52] indicated a departure from an axial ansatz and this was seen as supporting the picture of single particle orbits coupled to a triaxial core. This work concentrated on beta-stable nuclei, and recently similar results were found for an isotopic chain with nuclei outside of as well as within the valley of stability [53], suggesting a description by a triaxial projected shell model.

*Broken axiality seen in experimental data for even nuclei.*

*A* Thomas-Fermi mass evaluation [54] made for even nuclei indicated a rather weak sensitivity to axial symmetry breaking, but still in a number of randomly selected heavy nuclei a weak but clearly significant preference for triaxiality was predicted. Such a mass reduction in case of broken axial symmetry contradicts conclusions based on a liquid drop model [55] as published later. Electron scattering studies as well as Coulomb excitation and photon absorption to the lowest $2^+$-level have shown clear evidence for non-spherical shapes of many even nuclei [7, 19, 45, 56, 57], but axial symmetry breaking was discussed only rarely [51, 58 – 60], possibly because of a weak sensitivity which is in agreement to the Thomas-Fermi calculations [54]. This has led to a commonly used praxis to neglect axial symmetry breaking, and an intrinsic prolate axial deformation for most non-spherical even nuclei was concluded from the dominance of negative $Q_S$ values. Only in the Pt-Hg region as well as in a few other nearly magic isotopes positive $Q_S(2^+)$ were observed, indicating quasi-oblate shapes, i.e. negative $Q_i$. In a few cases complex Coulomb excitation studies [60, 61] have shown clear sensitivity to and evidence for broken axial symmetry; additional reorientation experiments allowed quadrupole measurements in nearly 200 even nuclei, some of which were supplemented by muonic X-ray studies [45]. In a triaxial rotational model [8] (TRM) the quadrupole quantities can be expressed by Eq. (6) and for comparison $Q_S$ from a cranking approximation [62] (CRA) is given in Eq. (7); both indicate proportionality to an intrinsic quadrupole moment $Q_i$. Using the knowledge of $\gamma$ one gets an expression for the lowest $2^+$ level and the traditional relation evolves for $\gamma \to 0$; $Q_0$ may be taken from the B(E2) to ground state (gs), as specified with Eq. (12).

$$\text{TRM:} \quad Q_S(2^+,1) = \frac{-2}{7} \cdot Q_i(\beta,\gamma) = \frac{-6\cos(3\gamma)}{7\sqrt{1+8\cos^2(3\gamma)}} \cdot Q_0(\beta) \underset{\gamma \to 0}{\to} \frac{-2}{7} \cdot Q_0(\beta) \quad (6)$$

$$\text{CRA:} \quad Q_S(2^+,1) = \frac{-2}{7} \cdot Q_i(\beta,\gamma) \simeq \frac{-4}{7}\cos(\gamma+60^o) \cdot Q_0(\beta) \underset{\gamma \to 0}{\to} \frac{-2}{7} \cdot Q_0(\beta) \quad (7).$$

Unfortunately the two approaches differ from each other by nearly a factor of 2, and this is pointed out again together with Figs. 3 and 4 and discussed there. The possibility of an experimental determination of **γ** was tested by various investigations [63-66] and from these studies as well as from other work on multiple Coulomb excitation [60, 61, 70, 71, 74, 75] the triaxiality parameter γ was found for many nuclei to differ significantly from zero; there is no evaluation available for **γ**-values such that the figure may contain differing values for the same nucleus. The data analysis will be improved considerably by involving quadrupole (rotation) invariants [61, 67]. In Fig. 3 the correlation of the observed $Q_0$ and cos(3γ) is displayed for more than 150 nuclei, for which relevant experimental data have been published. As is obvious from Eqs. (6) and (7) and depicted in Fig. 4 there is no unique and model independent way to derive $Q_i$ or $Q_0$ from observable quantities, respectively to relate $Q_i$ or $Q_s$ $(2^+,1)$ to $Q_0$. For the use in Fig. 3 the values for $Q_0$ were not derived from $Q_s$ with its large experimental uncertainties, using Eqs. (6) or (7), but rather calculated from compiled [68] experimental $B(E2,0 \to 2^+)$-values by using the axial limit in Eqs. (8, 9, 12); when the triaxiality parameter γ is already known one can regard $Q_0$ as quasi-observable besides it. The difference between the two models does not allow to derive $Q_i$ from $Q_s(2^+,1)$ unambiguously; in addition there is an important uncertainty from the fact that γ = 30° as well as $Q_0 = 0$ both lead to $Q_s = 0$.

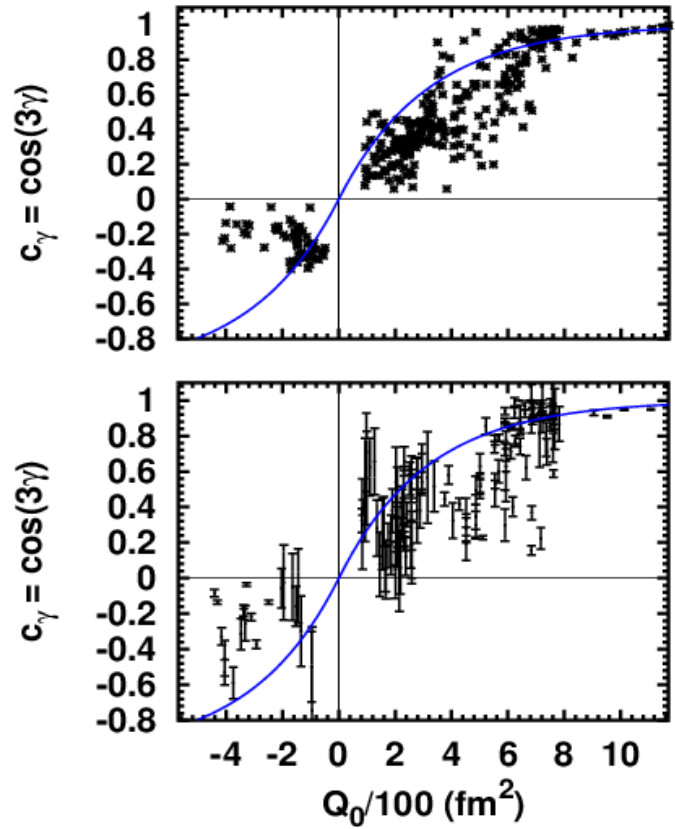

*Fig. 3: Correlation between $c_\gamma$=cos(3γ) and $Q_0$ in nearly 150 even nuclei with A>60, for which respective experimental data are available. In the upper panel results derived from level energy ratios $E(4^+)/E(2^+)$ as described in the text are plotted; the lower panel depicts cos(3γ) determined in various investigations quoted in the text and the bars correspond to experimental uncertainties. The x-axis in both panels corresponds to $Q_0$ derived from the $B(E2, 2 \to 0)$ and negative signs were taken from observations for Q.*
*The drawn blue curve serves as an eye-guide to compare the trend of the data to the one of the calculations shown in Fig. 5; it corresponds to sign(Q)·(1- exp(-|$Q_0$|/π)).*

Fig. 3 clearly indicates the trend of the axiality $c_\gamma$=cos(3γ) increasing with increasing $Q_0$; obviously well deformed nuclei show axial symmetry to a high degree. The few examples of oblate nuclei ($c_\gamma$=cos(3γ) < 0) do not allow for similar conclusions. The values for $c_\gamma$ as shown in the upper panel were derived by inserting published [68] values for $E(4^+)/E(2^+)$ into Eq. (4); here one respects the important role this energy ratio has played in a long lasting discussion on the rotational character of heavy nuclei, which also included values of $B(E2,0 \to 2^+)$ [69]. In the lower panel $c_\gamma$=cos(3γ) of such nuclei are depicted as published [60, 61, 63-65, 70-75]. Their large scatter indicates either possible systematic uncertainties not yet accounted for or the limitation of the approach presented here – which aims for a global treatment of apparent shell effects. Stressing mainly the breaking of axial symmetry one apparently extends the understanding of nuclear shape beyond a more local previous attempt [68].

*Theoretical considerations*

To depart from the axial limit and to account for γ ≠ 0 model calculations are needed and long ago an analytical treatment of a triaxial rotor was presented assuming rigid rotation of a non-axial body [8] and obtained an intrinsic quadrupole moment differing from the axial limit as shown in Eq.(8). A HFB-calculation has clearly indicated [76] the axial symmetry breaking also for nuclei often considered to be axially symmetric, if the HFB variation is performed properly after the projection out of the intrinsic frame. Related to that work is an attempt to derive a cranking approximation [62] to quadrupole properties of non-axial nuclei and respective results obtained by Eq.(9) to be compared here to Eq. (8) depicted in Fig. 4, which displays results from both equations for constant deformation $\beta$; differences of up to 15% are evident.

$$\text{TRM:} \quad 16\pi \cdot B(E2, 2^+_1 \to 0^+_{gs}) = \left(1 + \frac{1+2\cos^2(3\gamma)}{\sqrt{1+8\cos^2(3\gamma)}}\right) \cdot \frac{Q_0^2(\beta)}{2} \underset{\gamma \to 0}{\to} Q_0^2(\beta) \quad (8)$$

$$\text{CRA:} \quad 16\pi \cdot B(E2, 2^+_1 \to 0^+_{gs}) \approx \frac{\sqrt{3}}{\cos(30^\circ - \gamma)} \cdot \frac{Q_0^2(\beta)}{2} \underset{\gamma \to 0}{\to} Q_0^2(\beta) \quad (9).$$

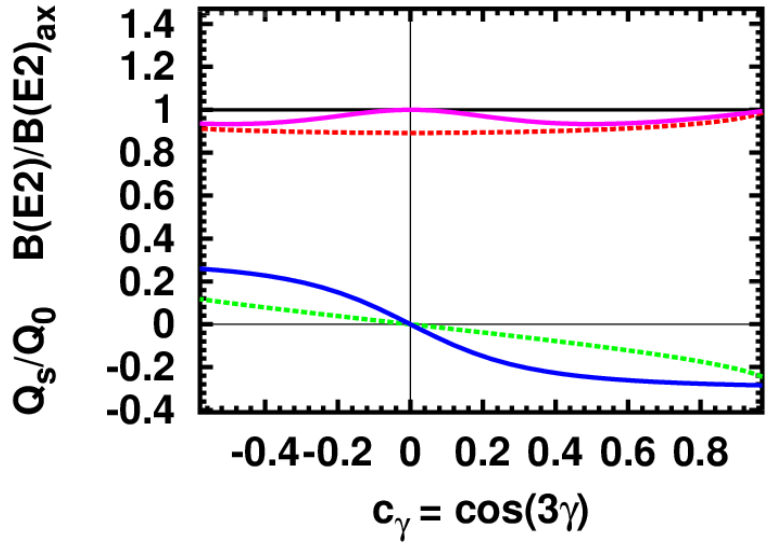

*Fig. 4: The lower set of curves depicts theoretically determined correlations between $c_\gamma = \cos(3\gamma)$ and $Q_s$, relative to the axial rigid rotor values; in the upper part of the figure the corresponding result is shown for B(E2). The drawn curves (blue and magenta) correspond to predictions from a rigid triaxial rotational model [8] and the dashed ones are from a cranking model approximation [62]. The deviations from the horizontal black lines indicate the error made when assuming a rigid axial shape.*

To overcome the ambiguities obvious from Fig. 4 more general considerations were proposed [67, 61] to show that a rotation invariant shape determination is possible, if all collective $2^+$-levels of the nucleus under study are included. A significant sensitivity to broken axial symmetry is indicated for the quadrupole transition strength between the two lowest $2^+$-states [8, 58, 60, 64, 74, 65, 75], as long as these are not clearly related to single particles. Although the rather complex experimental investigations on triaxiality have only been performed for a limited number of selected nuclei it is interesting to compare the correlation observed. Fig. 3 shows a recent calculation [77, 34] available for practically all heavy even nuclei between the neutron and proton drip lines. This study combines the Generator coordinate method (GCM) with a many particle calculation of constrained Hartree-Fock-Bogolyubov (CHFB) type. In Fig. 5 the correlation between **γ** and $Q_i$ is depicted for nuclei in the valley of stability as it results from this CHFB-GCM calculation, using Eq. (12). The calculation shows a very similar trend of the axiality increasing with $Q_i \approx Q_0$ as used in the experimental data plot; at variance to Fig. 3 the density of symbols is now of significance, as all the nuclei from a small band near stability were depicted; the clustering near 450 fm² belongs to lanthanide nuclei.

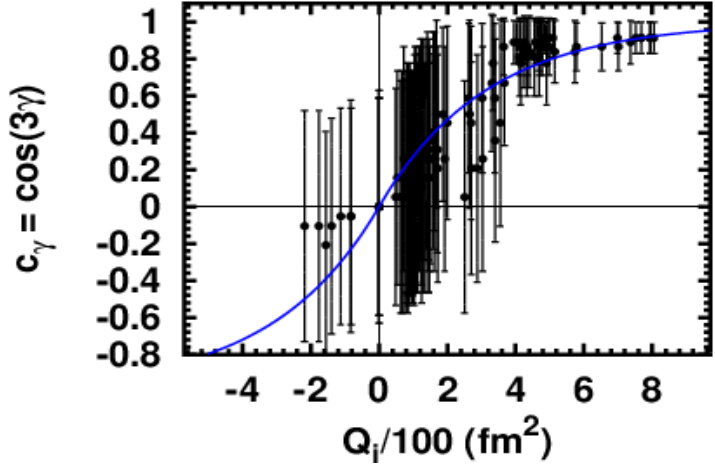

*Fig. 5: Correlation between cos(3γ) and $Q_i$ in about 130 even nuclei in the valley of stability with 60<A<240; the respective values are taken from a CHFB-GCM calculation [34]. The bar lengths represent the standard deviations in γ as given by these calculations, and the drawn blue line depicts the trend, it corresponds to sign(Q)·(1- exp(-|$Q_0$|/π)).*

The similarity between observation and calculation suggests to use the theoretical shape parameters also in predictions for the splitting of IVGDR shapes, which will be discussed later. The clustering at $Q_i \approx$ 200 fm$^2$ and cos(3γ) ≈ 0.2 seen in Fig. 5 is significant and plays an important role for the IVGDR evaluation by single Lorentzians versus the "triple Lorentzian" (TLO) parametrization [11, 26]. The CHFB-GCM-calculations [34] are performed in a triaxial oscillator basis and thus they result in the three frequencies $ω_x, ω_y, ω_z$, respectively their inverse, the half axes of an equivalent ellipsoid with volume V, $R_x, R_y, R_z$ which will be used in Eqs.(10, 11 & 13). The deformation parameters $β$ and $γ$ as explicitly defined and tabulated [34] are related to the axis lengths and an equivalent sphere radius $R_0$ by Eq. (3) of ref. [34]. One can also calculate them directly from $γ$ and $Q_i$ using Eqs. (13 & 19 -21) in the work on a concept of an equivalent ellipsoid [67] (CEE); this invokes a rather model independent view on the relation of the observables to geometrical quantities by assuming nothing but a homogeneous distribution of charge within the nuclear volume V:

$$\text{CEE:} \quad Q_i^2 = \frac{Z^2}{25}\left(2R_z^2 - R_x^2 - R_y^2\right)^2 - 3\left(R_x^2 - R_y^2\right)^2, \quad R_0^3 = R_x \cdot R_y \cdot R_z = \frac{3\pi}{4}V \quad (10)$$

$$\cos(3γ) \simeq \frac{\left(2R_z^2 - R_x^2 - R_y^2\right)^2 - 3\left(R_x^2 - R_y^2\right)^2}{\left(2R_z^2 - R_x^2 - R_y^2\right)^2 + 3\left(R_x^2 - R_y^2\right)^2} \quad (11)$$

Hence it is obvious that the observables $Q_i^2$ and $γ$ can be related directly to shape parameters in Cartesian coordinates without using the deformation parameter $β$, which has been related to spherical coordinates [19, 7]. As given in later work [67] a relation holds between $Q_i^2(β,γ)$ and B(E2), with the traditional one [7] for $Q_0^2 = Q_i^2(β,0)$ as limit:

$$Q_i^2(β,γ) = 16π \sum_k B(\text{E2}, 2^+_k \to 0^+_{gs}) \quad ; \quad Q_i^2(β,0) \simeq 16π \cdot B(\text{E2}, 2^+_1 \to 0^+_{gs}). \quad (12)$$

A survey [78] of experimental data for 101 even nuclei indicates that in 50% of the nuclei investigated more than 3% of quadrupole excitation strength from the ground state (gs) does not go to the first excited 2$^+$ level. A comparison of Eq's. (8 & 9) and (12) suggests to identify the higher components of the sum in Eq. (12) with the contribution from $γ > 0$ in Eqs. (8 & 9). Apparently Eq. (12) reveals the amount of quadrupole excitations aside from the first rotational level as measure of axiality breaking. Here this breaking has a very weak effect [78] whereas in the other paragraphs of this overview it is shown to cause significant changes. One related point has been studied experimentally some time ago [79]: All even isotopes of Pb show significantly strong excitation to 2$^+$-levels and these have non-zero spectroscopic quadrupole moments indicating a breaking of spherical symmetry – even if these are small.

Considering axial symmetry ($γ$ = 0°), quasi for the "other" 50% of heavy nuclei, the higher terms in the sum can be neglected and the resulting intrinsic electric quadrupole (λ=2) moment $Q_0$ can then be put in relation to the deformation parameter $β$, introduced [19] to describe the shape of even nuclei by spherical harmonics:

$$Q_0 \simeq \frac{3}{\sqrt{5\pi}} Z R^2 \beta (1+b\cdot\beta) \quad \text{and} \quad \beta \cong \frac{4}{3}\sqrt{\frac{\pi}{5}} \frac{R_z - R_{x,y}}{R_0} \cong 1.057 \frac{\Delta R}{R_0} \quad (13)$$

This relation (Eq. 13) disregards broken axiality, but it is widely used when electromagnetic data are related to calculated nuclear (mass) deformations often characterized by $\beta$. For years b ≈ 0.16 was used in many studies [7, 80], but a popular compilation [57] of B(E2)-values proposed b = 0 as an approximation. In the same paper b = 0.36 was used in a more detailed comparison to calculations assuming a zero hexadecapole moment. Besides the ambiguity in b there is another one: Several definitions proposed as deformation parameters in the literature [4, 19, 34, 57, 67] result in axis ratios, i.e. shapes, differing from each other especially for large $\beta$ or Q. Some of them – including the one [19] in widespread use – do not conserve the volume when departing from spherical symmetry, what may lead to unwanted contributions from compressional energy, and often these are neglected.

Another much studied aspect in the spectroscopy of even heavy nuclei, the angular momentum dependence of the yrast state energy may shed light on the importance of broken axial symmetry: Experiments for $^{238}$U [81] do not contradict to the axial rigid rotor approximation for B(E2), and with regard of these data one should not consider a spin dependent shape change (stretching) and disregard a variable moment of inertia (VMI), which was favoured at the time [cf. Ref. 16 in [81]]. With the rather small $\gamma$ = 8° from calculations [34] Eq. (4) results in a reasonable agreement to measured [81] data as shown in Fig. 6 which also compares to axial rigid rotation. Apparently the dependence on triaxiality proposed in Eq. (4) works at least as good as the VMI approach, which needs an additional fit parameter. In view of Fig. 4. the near independence of ground band B(E2)-values on $c_\gamma$, i.e. triaxiality $\gamma$, may well be considered understandable.

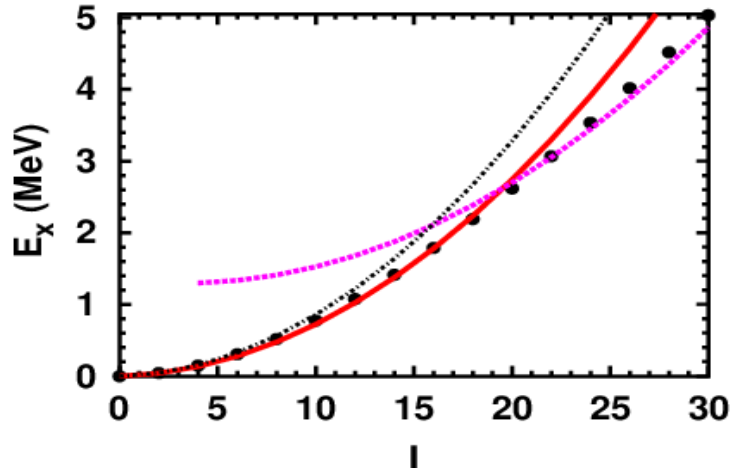

*Fig. 6 : The experimental spin dependence of the excitation energies in the yrast band of $^{238}$U (black dots • [81]) are compared to a rotational formula without (black dash-dot -•), and with dependence on $\gamma$ (red curve, cf. Eq. (4)), with $\gamma$ taken from a HFB calculation [34]. Obviously another band started at elevated excitation and angular momentum I = 4 is crossing near spin I = 20 (magenta dots). A ground band fit up to spin I = 18 resulted in a constant moment of inertia of $\Im$ =64ℏ²/MeV; this is 0.46 of $\Im_{rig}$ and 6.7 times $\Im_{irrot}$ [7]; for the other band an energy shift of 1.3MeV, the same $\Im$, and an increase of $\gamma$ from 8° to 15° in Eq.(4) were used, decreasing $c_\gamma$=cos(3$\gamma$) by 23%.*

Independent of the details in the definition of nuclear deformation the facts presented here make it obvious, that the justification for implying axiality has to be verified in each case, not only for those nuclei which "appear" axial or which are considered as such traditionally. And such a verification is especially needed before the frequently used [19] I·(I+1) - rule for the I-dependence of $E_{yr}$ is applied, which long ago was shown theoretically [82] to loose its validity in the absence of axial symmetry. Regarding Fig. 6 the influence of broken axiality should be considered the main cause for deviations from the I·(I+1) - rule instead of only assigning them to variations in the moments of inertia. The excursion from axiality presented here is likely to originate from nuclear shell structure and has to be distinguished from changes induced by rapid rotation discussed in the last paragraphs of ref. [19].

### IV. Triple splitting of IVGDR's and photon strength

In nuclear astrophysics as well as for plans concerning the transmutation of radioactive waste the radiative capture especially of neutrons plays an important role. Respective cross sections are based on statistical model calculations and the reliability of predictions depends not only on the proper characterization of the input channel by e.g. an optical potential, but even more strongly on the details determining the decay of the intermediately formed compound nucleus. Here the strength of its electromagnetic decay is of importance as well as the open phase space in the final nucleus, i.e. the density of levels reached by the first photon emitted. The importance of broken axial symmetry for the latter has already been demonstrated in section II, but also radiative widths play an important role for the de-excitation [83, 84] following the capture process. Here the electromagnetic strength function has to be known and the large cross section for the photo-disintegration of nuclei has been recognized as outstanding means to determine it experimentally – not much after it was studied as one of the first nuclear reactions [85]. The resonant structures observed in heavy nuclei were identified as a manifestation of collective vibrational dipole and quadrupole excitation mode [86, 87] and its coupling to other degrees of freedom has been much discussed [19, 80, 88, 89]. Especially the splitting of the IVGDR in lanthanide and actinide nuclei – as obvious in the experimental data [90, 91] – was discussed as manifestation of their deformation since long-time [19, 88, 93]. The role of axial symmetry breaking was worked out by a respective reanalysis of photo-neutron data as performed [11, 27, 95, 94] recently using the compilation EXFOR [96]; these data stem from old experiments, which have been repeated very rarely only.

*Photon absorption cross section*

From quantum electrodynamics considerations like causality, analyticity and dispersion relations the Thomson scattering cross section can be generalized to shorter wavelength photons interacting with nuclei of mass number A. This leads to the Gell-Mann, Goldberger, Thirring (GGT) sum rule [97]:

$$\int_0^{m_\pi c^2} \sigma\, dE \approx 2\pi^2 \frac{\alpha \hbar^2}{m_N}\left[\frac{ZN}{A}+\frac{A}{10}\right] = 5.97\left[\frac{ZN}{A}+\frac{A}{10}\right] MeV fm^2 \qquad A = Z+N \qquad (14)$$

Here $m_N$ and $m_\pi$ stand for the mass of nucleon and pion, respectively. The first term in the sum is the "*classical sum rule*" and the second "*contains all of the mesonic effects*" and is assumed [98] to be accurate within 30%. It was approximated by assuming "*that a photon of extremely large energy interacts with the nucleus as a system of free nucleons*", and a weak correlation to hadronic shadowing was investigated [98]. The nuclear absorption of photons with energies above $m_\pi c^2$ can be disregarded [97] concerning the possible contributions to lower energy. Apparently [11] the absorption above 40 MeV mainly corresponds to the quasi-deuteron mode [99] and its integral is close to the second term in Eq.(14). The good agreement in the region below 20 MeV indicates that this first term describes the IVGDR, i.e. the isovector nuclear electric dipole mode. A large excess above the Thomas-Reiche-Kuhn [100] (TRK) sum rule has been reported [91], whereas allowing for triaxiality such a disagreement becomes negligible [11, 94].

*Giant resonance energies*

The first theoretical descriptions for the oscillation of protons against neutrons described medium mass nuclei [86] and the very heavy ones [87] reasonably well. By using the concept of the droplet model these two approaches were unified and IVGDR centroid energies $E_0(Z,A)$ are well predicted [101] in the range 60<A<240. As nuclear radius the charge radii from the CHFB-GCM-calculations [34] were taken [11] and a symmetry energy J=32.7 MeV and a surface stiffness Q=29.2 MeV from the finite range droplet model [102] were used. Only one additional parameter, an effective nucleon mass $m_{eff}$ = 800 MeV, had to be adjusted to give an overall fit to the IVGDR data for a large number of nuclei with 70<A<240. The predictive power of the parametrization proposed for the absorption $\sigma_{ab}$ was studied by comparing the sum of three Lorentzians to experimental data [103, 94, 26, 95, 11]

$$\sigma_{ab}(E) = 5.97 \frac{ZN}{A}\frac{2}{3\pi}\sum_{k=1}^{3}\frac{E^2 \Gamma_k}{\left(E_k^2-E^2\right)^2 + E^2 \Gamma_k^2} fm^2 ; E=E_\gamma \qquad (15)$$

to properly adjusted [104, 105] experimental IVGDR data from the literature [96], and a clear advantage of using three Lorentzian components (TLO) was demonstrated. Here the $E_k$ and $\Gamma_k$ are derived from the centroid predictions by regarding the broken axiality as presented above. Hence, the main term of eq. (14), the classical sum rule, is divided equally into three portions. The deformation induced shift of the three energies $E_k$ versus the centroid energy $E_0$ is obtained from Eq. (3) of a recent CHFB-GCM calculation [34, 77]. A satisfactory description of the IVGDR shape especially near the maximum of the cross section is obtained [11] only by using 3 poles at energies $E_k$. The height of the low energy slope, of importance for the radiative capture cross section, also agrees to experimental data.

*Giant resonance widths*

The apparent shape and width $\Gamma_{app}$ of the IVGDR in heavy nuclei is strongly influenced by a significant fragmentation predicted theoretically [106]. It was demonstrated [21] that this is very likely related to a splitting induced by the nuclear shape and it was shown [11, 94, 107] that broken axiality is responsible for the actually observed resonance form, in contrast to statements made in a recent paper on a more microscopic calculation [108] based on a random phase approximation. Allowing for that triple split a simultaneous overall fit could be performed [11] for more than 20 isotopes simultaneously and only one global width parameter $c_w = 0.045$ was needed when using a predicted [109] energy dependence $\Gamma_k = c_W \cdot E_k^{1.6}$. This resulted in slightly different widths of the three poles in each isotope in dependence of their energy; as apparent from the data this procedure worked for all nuclei studied [11]. In accordance to a precision study [104] confirmed by results [105, 94] from the radiation source ELBE, the photo-neutron data [96] of the Saclay group [93, 110, 111] had to be reduced by 12%. This change, an accounting for the rather large dispersion of quasi-monochromatic photon beams and the application of the TLO approach allowed a consistent description of the IVGDR in more or less all heavy nuclei with one global width parameter $c_W$ and hence no explicit A-dependence as well as no difference to the TRK sum rule. And it also made an application of a photon energy dependence of radiative widths [112] to electric giant resonances superfluous, which was often made [113] for various nuclei. In addition, a triple split was already indicated for magnetic dipole resonances [114] as observed by electron scattering.

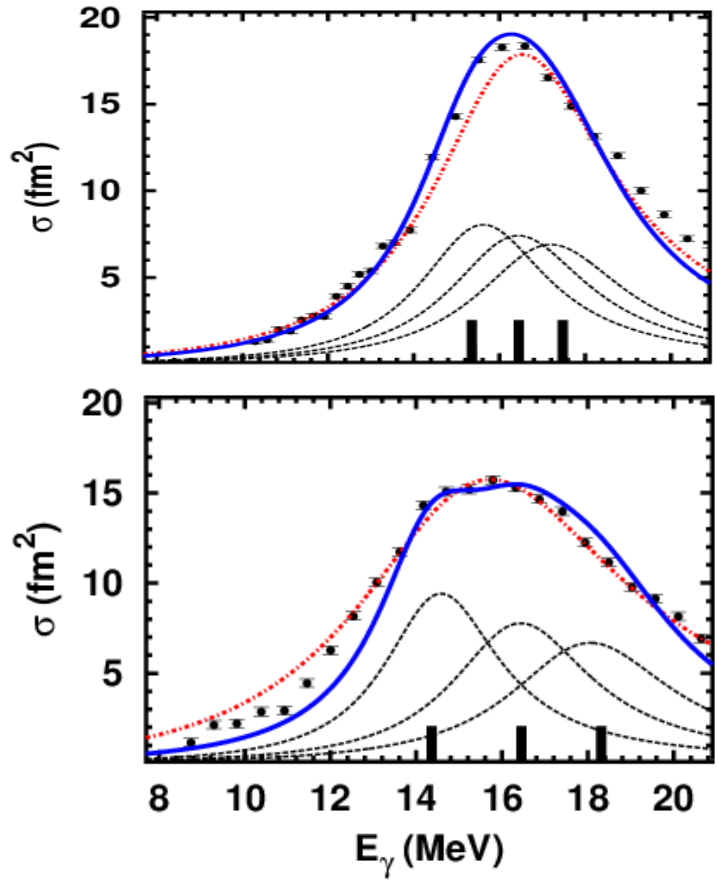

*Fig 7 : Cross section of photo-neutron production [111] on $^{94}$Mo (top) and $^{100}$Mo (bottom) in comparison to the calculated sum for three Lorentzians (TLO) for the GDR (blue drawn line); the shapes and centroids of the three components are indicated in black and correspond to (β ≈ 0.16, γ ≈ 28°) and (β ≈ 0.23, γ ≈ 26°), respectively. The red dash-dotted line depicts the result of a 3-parameter fit to each data set by a single Lorentzian (SLO) [91]. The TLO approach considers the extra cross section outside of the IVGDR as due to other modes (e.g. pygmy, M1, E2) which cannot be covered by an electric dipole description of the IVGDR.*

*IVGDR widths and tails*

Fig. 7 shows results for the IVGDR shapes calculated for two Mo isotopes; predictions for these were already published [10, 94] previous to the CHFB-GCM-calculations, and the analysis was newly repeated using it. From the comparison between the results for $^{94}$Mo to $^{100}$Mo the main advantages of working with three poles (TLO) become obvious: The IVGDR in the two isotopes are well reproduced with the same parameters for position and width, in contrast to separate fits [92] with a single Lorentzian (SLO), producing $\Gamma_{94}$ = 5.12 MeV and $\Gamma_{100}$ = 7.68 MeV for the two isotopes. The SLO fits [20, 91, 92] as also depicted in Fig. 5 give a clear hint to the reason for a need of 6 parameters for the 2 isotopes: Not only that it neglects the deformation induced split – which is not directly obvious due to the middle pole hiding a possible minimum; the fit also includes strength in the tail part of the cross section with the IVGDR seemingly widened by the inclusion of the three poles in one Lorentzian. Near 20 MeV as well as between 9 and 12 MeV quadrupolar strength may have been observed in various heavy nuclei [11, 103, 115, 116] and earlier work quoted [11] and is expected to also appear in Mo isotopes [94, 117].

Similar statements can be made for the odd A nucleus $^{181}$Ta, in which again the quadrupole giant resonance lies in energy a few MeV above the IVGDR. As presented in Fig. 8, the IVGDR energy, its yield and the width are well predicted, with the triaxial deformation well accounted for by the predicted values for $\beta$ and $\gamma$ [34]. And the fit made for RIPL-3 [20, 92] misinterprets the shape due to the disregard of triaxiality. The resulting strong variation of the spreading width from $E_1$ = 12 MeV to $E_2$ = 15 MeV seems impossible to be explained. In addition, this 2-pole fit [91, 92] sees 28 % excess over the TRK sum-rule and also this point lacks theoretical explanation.

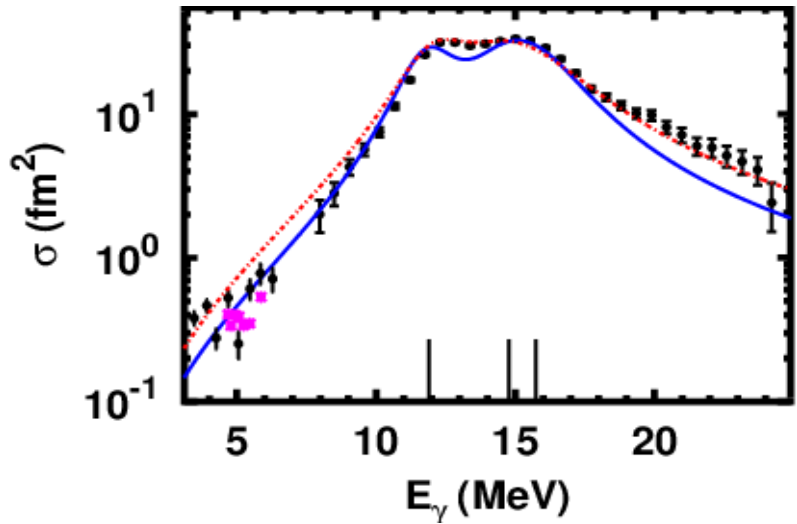

*Fig. 8: Photon absorption cross section derived for $^{181}$Ta from photon scattering [118] (black dots, below 7 MeV, modified by 0.7, and magenta \*-symbols[119]) and from photo-neutron production (black dots, [110]) in comparison to TLO (described in text) and hence to the TRK sum rule. Also shown is a double Lorentzian fit curve from RIPL-3 [20, 92] (red dash-dotted line; adjusted in height to re-normalized data [104, 105, 94]. It results in very different widths for the two peaks of $\Gamma_1$ = 2.5 MeV and $\Gamma_2$ = 4.4 MeV and consequently in nearly 30 % overshoot over the TRK sum in contrast to TLO for which there is no need for a strong deviation of the energy integrated strength from it. The observed excess below 4 and above 18 MeV is likely to originate from non-E1 modes [11].*

The IVGDR parametrization (TLO) [11], as reviewed here, makes a detailed use of the predictions for the non-axial deformation of heavy nuclei with A>60 as obtained recently by CHFB-GCM-calculations [34], mentioned above. This leads to an agreement of the integrated resonance strengths to sum-rule predictions and to a rather straightforward description of the spreading widths by one global parameter, the proportionality factor in the power law for the GDR pole energy dependence of the reonance widths – with the exponent fixed already theoretically [109]. When TLO is extended to lower energy, which is of importance for radiative capture, it fits or under-predicts earlier observations [11] leaving room for absorption channels adding to the IVGDR dipole mode.

*Radiative neutron capture*

The TLO approach [94, 26, 11] can be combined to the quasi parameter-free ansatz for the level densities [28] described above to predict yields for the capture of thermal neutrons with subsequent photon emission. For comparison a survey of various experiments was made [120 - 122] and presented radiative neutron capture cross section data for about 130 even target nuclei, Maxwellian averaged at kT = 30 keV. A prediction the cross section for forming a compound nucleus by radiative capture is obtained by multiplying the geometrical cross section to a convolution of level density and radiative strength; this can be performed easily when a statistical decay [123, 124] of the final nucleus by gamma rays is assumed. A surprisingly good agreement of such a procedure to observation, i.e. to the averaged radiative cross sections, has been reached [28, 29] and equivalent results for the Maxwellian averaged neutron capture cross section are are depicted in Fig. 9. Here the photon strength is derived from the parameters well describing the IVGDR pole energies by $m_{eff}$ and a global fit of their width together with using the classical dipole sum rule. The other ingredient, the level densities, depend mainly on the shell correction needed – and this is already determined from the fit to nuclear masses. As shown already in Fig. 1 the respective level densities agree well to observed neutron capture resonance spacings. The good accordance depicted in Fig. 9 indicates the value of the proposed approach.

Fig. 9 also shows the possible effect of various electromagnetic excitation modes in heavy nuclei at low excitation energy and not related to the IVGDR tail. The systematic study [11, 26] of it should be a good basis to test the influence of these extra modes, some of which were proposed there recently.

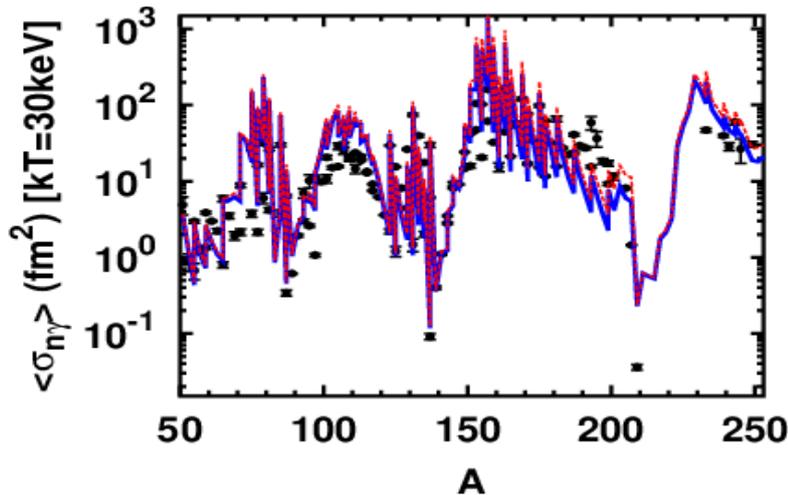

*Fig. 9: Plot of the Maxwellian averaged neutron capture cross section shown as black dots [120, 121, 122] for even-even nuclei together with the TLO ansatz using three Lorentzians combined to the discussed parameter free level density prediction (drawn curve in blue) versus A in the valley of stability. The difference to the upper dashed curves (in red) depicts the gain in cross section due to photon strength not related to the IVGDR, compiled recently [11] from various measurements.*

The combination of the low energy slope resulting from TLO to the parameter free prediction of level densities (ch. II) leads to a surprisingly good description for the cross section of radiative neutron capture. We thus consider this ansatz a very good starting point for network calculations in the field of nuclear astrophysics and especially for the element synthesis in the s- and r-processes. It is also applicable for the numerical simulation of nuclear power systems and the transmutation of radioactive waste. Together with the significant improvement in the description of IVGDR shapes – including their positions and widths – the agreement seen in Fig. 7 constitutes another support of admitting the breaking of axial symmetry in heavy nuclei already in the valley of stability. Here it should be pointed out that the rather large deviation of prediction from observation for A > 200 needs further investigation especially of the shell correction entering Eqs. (1) and (2). Also the impact of magnetic dipole strength [125-127] should be mentioned as well as the demonstration of minor impact of a stochastic variation of the deformation parameters by instantaneous shape sampling (ISS) on the shape of the IVGDR's in various nuclei [11, 94, 128]. A number of publications from the Dresden-Rossendorf group [11, 26, 27, 30, 94, 95, 115, 116, 129, 130] extended the work on IVGDR shapes and finally the atomic number range to 70< A < 240 was covered showing that in these heavy nuclei assumption of spherical or axial symmetry is disfavored to reach a globally valid description of isovector giant dipole resonances.

## V. Conclusions

This overview paper has presented the impact of broken axial symmetry on predictions for various observables from different fields of nuclear physics. A prediction of level densities and photon strength functions based on a breaking of axial and spherical symmetries can be constructed by only using global parameters already known from the study of other observables – in contrast to a number of approaches reviewed recently [20] usually fitting several parameters locally to observed IVGDR-shapes and to level spacing measured *e.g.* in thermal neutron capture. Turning away from the quite common preference for axiality of heavy nuclei in the valley of stability allows a quasi parameter-free prediction of these observations and in addition a good description for Maxwellian averages of neutron capture cross sections in a wide range of A; these are of importance for nuclear astrophysics. The breaking of axial symmetry was emphasized by CHFB-GCM calculations [34] and such a symmetry breaking also was applied – nearly simultaneously – to an analysis [94] of IVGDR data for isotopes of Mo. Recent more global extensions of that work [11] came to the same conclusion for an extended number of nuclei including nuclides near magic shells as well as well deformed ones. Both publications demonstrated an only sooth change of the IVGDR's spreading widths with no explicit A-dependence and it added interesting information on their low energy tails. These results are in accord to the analysis of nuclear spectroscopic observations made some time ago and a remarkable fact evolved: The axiality is nearly perfect for large $|Q_i|$ with $|c_\gamma| = |\cos(3\gamma)| \cong 1$ and is clearly broken for smaller $|Q_i|$ and $c_\gamma \cong 0$. It appears as if $c_\gamma$ can be considered quasi as general scaling parameter for the dependence of various observables on nuclear axiality and its breaking. From the experimental data presented here and the quoted work on theoretical considerations one can undoubtedly recommend to no longer assume *ad hoc* that quasi all heavy nuclei are either spherical, axially symmetric or have at the very most a dynamical triaxiality only. In view of the facts discussed in this overview paper the assumption of static triaxiality seems clearly to be favored over an only dynamical breaking of axial symmetry already for nuclei in the valley of stability. This demonstrates a certain similarity to an observation made for the breaking of symmetries in solids[131,132] as induced by quantum effects.

This work was supported within the 7th Framework Program of the European Commission through Fission 2013 — CHANDA Work Package 4 (605203).
Intense discussions within this project and with other colleagues, especially with Hans Feldmeier, Ralph Massarczyk, Ronald Schwengner, Julian Srebrny and Hermann Wolter are gratefully acknowledged.